\documentclass[final,authoryear,5p,times,twocolumn]{elsarticle}
\pdfoutput=1

\usepackage{graphicx}

\usepackage{amssymb}

\usepackage[pdftex,pdfpagemode={UseOutlines},bookmarks,bookmarksopen,colorlinks,linkcolor={blue},citecolor={green},urlcolor={red}]{hyperref}
\usepackage{hypernat}

\journal{Astronomy \& Computing}

\usepackage{upquote}

\usepackage{upgreek}

\usepackage{color}

\newcommand*\secref[1]{Sect.~\ref{#1}}

\begin{document}

\begin{frontmatter}



\title{ORAC-DR: A generic data reduction pipeline infrastructure}


\author[jac]{Tim Jenness\corref{cor1}\fnref{timj}}
\ead{tjenness@cornell.edu}
\author[jac]{Frossie Economou\fnref{fe}}

\cortext[cor1]{Corresponding author}
\fntext[timj]{Present address: Department of Astronomy, Cornell University, Ithaca,
  NY 14853, USA}
\fntext[fe]{Present address: LSST Project Office, 933 N.\ Cherry Ave, Tucson, AZ 85721, USA}

\address[jac]{Joint Astronomy Centre, 660 N.\ A`oh\=ok\=u Place, Hilo, HI
  96720, USA}

\begin{abstract}

  ORAC-DR is a general purpose data reduction pipeline system designed
  to be instrument and observatory agnostic. The pipeline works with
  instruments as varied as infrared integral field units, imaging
  arrays and spectrographs, and sub-millimeter heterodyne arrays \&
  continuum cameras. This paper describes the architecture of the
  pipeline system and the implementation of the core
  infrastructure. We finish by discussing the lessons learned since
  the initial deployment of the pipeline system in the late 1990s.

\end{abstract}


\end{frontmatter}


\newcommand{\mnras}{MNRAS}
\newcommand{\aap}{A\&A}
\newcommand{\aaps}{A\&AS}
\newcommand{\pasp}{PASP}
\newcommand{\apj}{ApJ}
\newcommand{\apjs}{ApJS}
\newcommand{\qjras}{QJRAS}
\newcommand{\an}{Astron.\ Nach.}
\newcommand{\ijimw}{Int.\ J.\ Infrared \& Millimeter Waves}
\newcommand{\procspie}{Proc.\ SPIE}
\newcommand{\aspconf}{ASP Conf. Ser.}


\newcommand{\recipe}{\emph{Recipe}}
\newcommand{\recipes}{\emph{Recipes}}
\newcommand{\primitive}{\emph{Primitive}}
\newcommand{\primitives}{\emph{Primitives}}
\newcommand{\Frame}{\emph{Frame}}
\newcommand{\Group}{\emph{Group}}
\newcommand{\Index}{\emph{index}}

\newcommand{\oracdr}{\textsc{orac-dr}}
\newcommand{\cgsdr}{\textsc{cgs}{\footnotesize 4}\textsc{dr}}

\newcommand{\ascl}[1]{\href{http://www.ascl.net/#1}{ascl:#1}}


\section{Introduction}

In the early 1990s each instrument delivered to the United Kingdom
Infrared Telescope (UKIRT) and the James Clerk Maxwell Telescope (JCMT) came
with its own distinct data reduction system that reused very little
code from previous instruments. In part this was due to the rapid
change in hardware and software technologies during the period, but it
was also driven by the instrument projects being delivered
by independent project teams with no standardisation requirements
being imposed by the observatory. The observatories were required to
support the delivered code and as operations budgets shrank the need
to use a single infrastructure became more apparent.

\cgsdr\
\citep[][\ascl{1406.013}]{1992ASPC...25..479S,1996ASPC...87..223D} was
the archetypal instrument-specific on-line data reduction system at
UKIRT. The move from VMS to UNIX in the acquisition environment coupled
with plans for rapid instrument development of UFTI
\citep{2003SPIE.4841..901R}, MICHELLE \citep{1993ASPC...41..401G} and
UIST \citep{2004SPIE.5492.1160R}, led to a decision to revamp the
pipeline infrastructure at UKIRT \citep{1998ASPC..145..196E}. In the
same time period the SCUBA instrument \citep{1999MNRAS.303..659H} was
being delivered to the JCMT. SCUBA had an on-line data reduction
system developed on VMS that was difficult to modify and ultimately
was capable solely of simple quick-look functionality. There was no explicit
data reduction pipeline and this provided the opportunity to develop a
truly instrument agnostic pipeline capable of supporting different
imaging modes and wavelength regimes.

The Observatory Reduction and Acquisition Control Data Reduction pipeline
\citep[\oracdr;][\ascl{1310.001}]{1999ASPC..172...11E,2008AN....329..295C} was
the resulting system. In the sections that follow we present an
overview of the architectural design and then describe the pipeline
implementation. We finish by detailing lessons learned during the
lifetime of the project.

\section{Architecture}

The general architecture of the \oracdr\ system has been described
elsewhere \citep{1999ASPC..172...11E,2008AN....329..295C}. To
summarize, the system is split into discrete units with well-defined
interfaces. The recipes define the processing steps that are required
using abstract language and no obvious software code. These recipes
are expanded into executable code by a parser and this code is
executed with the current state of the input data file objects and
calibration system. The recipes call out to external
packages\footnote{These are known as ``algorithm engines'' in the
 ORAC-DR documentation.} using a standardized calling interface and it is these
applications that contain the detailed knowledge of how to process pixel
data. In all the currently supported instruments the external algorithm
code is from the Starlink software collection
\citep[][\ascl{1110.012}]{2014ASPC..485..391C} and uses the ADAM
messaging system \citep{1992ASPC...25..126A}, but this is not
required by the \oracdr\ design. There was a deliberate decision to
separate the core pipeline functionality from the high-performance
data processing applications so that one single application
infrastructure was not locked in.

\begin{figure*}
\includegraphics[width=\textwidth]{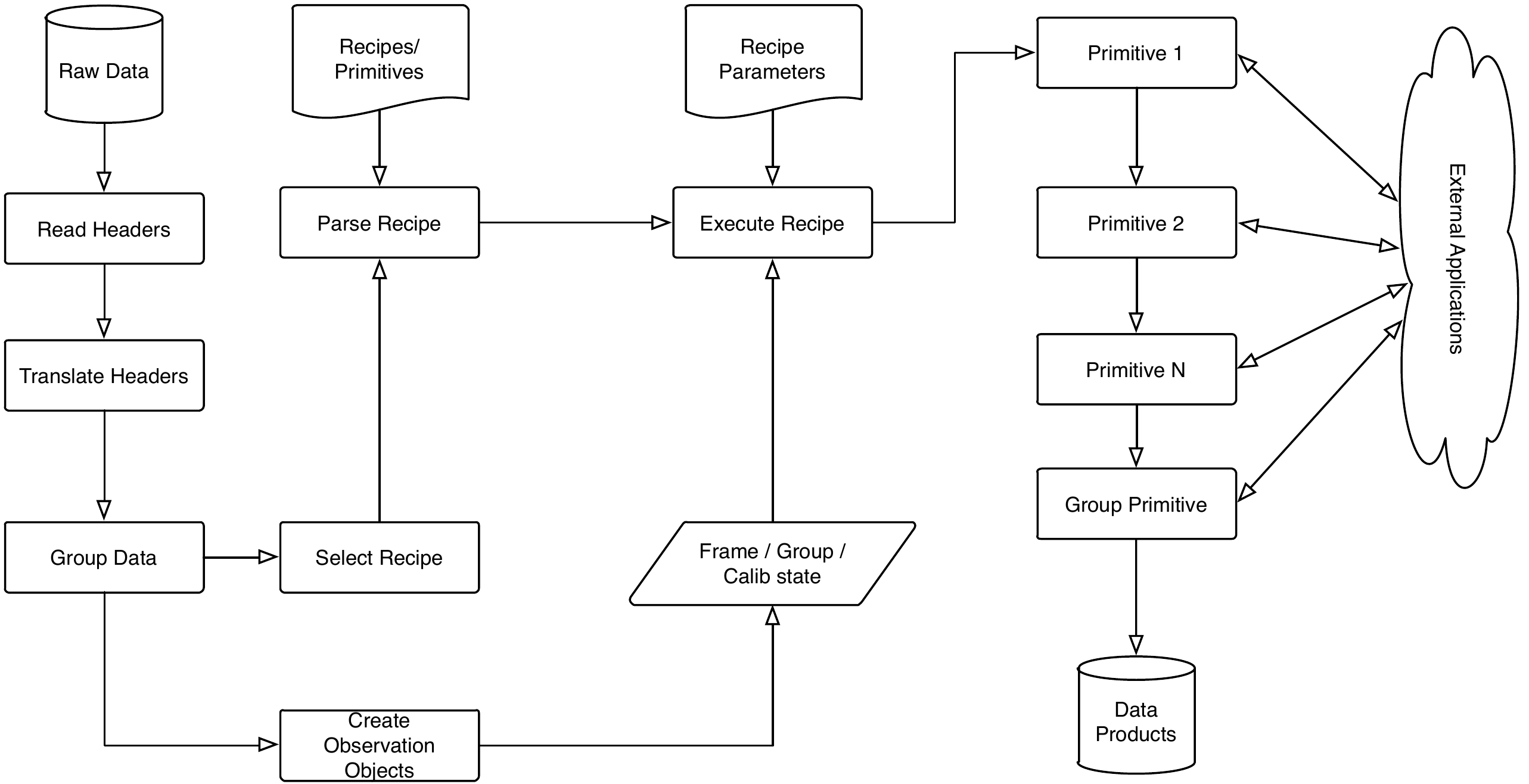}
\caption{Outline of the control flow in ORAC-DR for a single
  observation. For multiple observations the pipeline will either
  check for more data at the end of the \recipe\ execution (on-line
  mode) or read all files and do group assignments before looping
  over groups (batch mode).}
\label{fig:flow}
\end{figure*}

A key part of the architecture is that the pipeline can function
entirely in a data-driven manner. All information required to reduce
the data correctly must be available in the metadata of the input data
files. This requires a systems engineering approach to observatory
operations where the metadata are treated as equals to the science
pixel data \citep[see e.g.,][for an overview of the JCMT and UKIRT
approach]{2011tfa..confE..42J} and all observing modes are designed
with observation preparation and data reduction in mind. An overview
of the pipeline process is show in Fig.~\ref{fig:flow}.

\section{Implementation}

In this section we discuss the core components of the pipeline
infrastructure. The algorithms themselves are pluggable parts of the
architecture and are not considered further. The only requirement
being that the algorithm code must be callable either directly from
Perl or over a messaging interface supported by Perl.

\subsection{Data Detection}

The first step in reducing data is determining which data should be
processed. \oracdr\ separates data detection from pipeline processing,
allowing for a number of different schemes for locating files. In
on-line mode the pipeline is set up to assume an incremental delivery
of data throughout the period the pipeline is running. Here we
describe the most commonly-used options.

\subsubsection{Flag files}

The initial default scheme was to check whether a new file with the
expected naming convention had appeared on disk. Whilst this can work
if the appearance of the data file is instantaneous (for example, it
is written to a temporary location and then renamed), it is all too
easy to attempt to read a file that is being written to. Modifying
legacy acquisition systems to do atomic file renames proved to be
difficult and instead a ``flag'' file system was used.

A flag file was historically a zero-length file created as soon as the
observation was completed and the raw data file was closed. The
pipeline would look for the appearance of the flag file (it would be
able to use a heuristic to know the name of the file in advance and
also look a few ahead in case the acquisition system had crashed) and
use that to trigger processing of the primary data file.

As more complex instruments arrived capable of writing multiple files
for a single observation (either in parallel
\citep[SCUBA-2;][]{2013MNRAS.430.2513H} or sequentially
\citep[ACSIS;][]{2009MNRAS.399.1026B}) the flag system was
modified to allow the pipeline to monitor a single flag file but
storing the names of the relevant data files inside the file (one file
per line). For the instruments writing files sequentially the pipeline
is able to determine the new files that have been added to the file
since the previous check.

Historically synchronization delays over NFS mounts caused
difficulties when the flag file would appear but the actual data file
had not yet appeared to the NFS client computer, but on modern systems
this behavior no longer occurs. Modern file event notification schemes
(such as \texttt{inotify} on Linux) do not generally help with the
data detection problem since, in the current setup, the data reduction
pipelines always  mount the data disks from the acquisition computer
over NFS. A more robust solution is to implement a publish/subscribe
system whereby the pipeline monitors the acquisition computers for new
data. Such a scheme is discussed in the next section.

\subsubsection{Parameter monitoring}

The SCUBA-2 quick look pipeline \citep{2005ASPC..347..585G} had a
requirement to be able to detect files taken at a rate of
approximately 1\,Hz for stare observations. This was impractical using
a single-threaded data detection system embedded in the pipeline process and
using the file system. Therefore, for SCUBA-2 quick-look processing the
pipeline uses a separate process that continually monitors the
four data acquisition computers using the DRAMA messaging system
\citep{1995SPIE.2479...62B}. When all four sub-arrays indicate that a
matched data set is available the monitored data are written to disk
and a flag file created. Since these data are ephemeral there is a
slight change to flag file behavior in that the pipeline will take
ownership of data it finds by renaming the flag file. If that happens
the pipeline will be responsible for cleaning up; whereas if the
pipeline does not handle the data before the next quick look image
arrives the gathering process will remove the flag file and delete the
data before making the new data available.

\subsection{File format conversion}

Once files have been found they are first sent to the format
conversion library. The instrument infrastructure defines what the
external format of each file is expected to be and also the internal format
expected by the reduction system. The format conversion system knows
how to convert the files to the necessary form. This does not always
involve a change in low level format (such as FITS to NDF) but can
handle changes to instrument acquisition systems such as converting
HDS files spread across header and exposure files into a single HDS
container matching the modern UKIRT layout.

\subsection{Recipe Parser}

A \recipe\ is the top-level view of the data processing steps
required to reduce some data. The requirements were that the recipe
should be easily editable by an instrument scientist without having to
understand the code, the \recipe\ should be easily understandable by
using plain language, and it should be possible to reorganize steps
easily. Furthermore, there was a need to allow \recipes\ to be edited
``on the fly'' without having to restart the pipeline. The next data file
to be picked up would be processed using the modified version of the
\recipe\ and this is very important during instrument commissioning. An
example, simplified, imaging \recipe\ is shown in Fig.\
\ref{fig:recipe}. Each of these steps can be given parameters to
modify their behavior. The expectation was that these \recipes\ would
be loadable into a Recipe Editor GUI tool, although such a tool was
never implemented.

\begin{figure}
{
\small
\begin{verbatim}
_SUBTRACT_DARK_
_DIVIDE_BY_FLAT_
_BIAS_CORRECT_GROUP_
_APPLY_DISTORTION_TRANSFORMATION_
_GENERATE_OFFSETS_JITTER_
_MAKE_MOSAIC_ FILLBAD=1 RESAMPLE=1
\end{verbatim}
}
\caption{A simplified imaging \recipe. Note that the individual steps
  make sense scientifically and it is clear how to change the order or
  remove steps. The \texttt{\_MAKE\_MOSAIC\_} step includes override
  parameters.}
\label{fig:recipe}
\end{figure}

Each of the steps in a \recipe\ is known as a
\primitive. The \primitives\ contain the Perl source code and can
themselves call other \primitives\ if required.
The parser's core job is to read the \recipe, replace the mentions of \primitives\
with subroutine calls to the source code for that primitive. For each
\primitive\ the parser keeps a cache containing the compiled form of
the \primitive\ as a code reference, the modification time associated
with the \primitive\ source file when it was last read, and the full
text of the \primitive\ for debugging purposes. Whenever a \primitive\
code reference is about to be executed the modification time is
checked to decide whether the \primitive\ needs to be re-read.

The parser is also responsible for adding additional code at the
start of the \primitive\ to allow it to integrate into the general
pipeline infrastructure. This code includes:

\begin{itemize}

\item Handling of state objects that are passed through the subroutine
  argument stack and parsing of parameters passed to the \primitive\
  by the caller. These arguments are designed not be language-specific
  and use a simple \texttt{KEYWORD=VALUE} syntax
  and can not be handled directly by the Perl interpreter.

\item Trapping for \primitive\ call recursion.

\item Debugging information
such as timers to allow profile information be
collected, and entry and exit log messages to indicate exactly when
a routine is in use.

\item Callbacks to GUI code to indicate which \primitive\ is
currently active.

\item Configuring the logging system so that all messages appearing
 will be associated with the correct primitive when they are written
to the history blocks (see \secref{sec:prov} for details).

\end{itemize}

The design is such that adding new code to the entry and exit of each
\primitive\ can be done in a few lines with little overhead. In
particular, use is made of the \verb|#line| directive in Perl that
allows for the line number to be manipulated such that error messages
reflect the line number in the original \primitive\ and not the line
number in the expanded \primitive.

Calling external packages is a very common occurrence and is also where
most of the time is spent during \recipe\ execution. In order to
minimize repetitive coding for error conditions and to allow for profiling, calls to
external packages are surrounded by code to automatically handle these
conditions. This allows the programmer to focus on the \recipe\ logic
and not have to understand all the failure modes for a particular
package.\footnote{The \texttt{oracdr\_parse\_recipe} command can be run
to provide a complete translation of a \recipe.}  The parser is
designed such that if a particular error code is important (for
example there might be an error code indicating that a failure was due
to there being too few stars in the image) then the automated error
handling is changed if the \primitive\ writer is explicitly asking to
check the return value from the external application.

\subsection{Recipe Parameters}

The general behavior of a recipe can be controlled by editing it and
adjusting the parameters passed to the \primitives. A much more
flexible scheme is available which allows the person running the
pipeline to specify a \recipe\ configuration file that can be used to
control the behavior of \recipe\ selection and how a \recipe\ behaves.

The configuration file is a text file written in the INI
format. Although it is possible for the \recipe\ to be specified on
the command-line that \recipe\ would be used for all the files being
reduced in the same batch and this is not an efficient way to
permanently change the \recipe\ name. Changing the file header is not
always possible so the configuration file can be written to allow
per-object selection of \recipes. For example,

\begin{quote}
\begin{verbatim}
[RECIPES_SCIENCE]
OBJECT1=REDUCE_SCIENCE
OBJECT2=REDUCE_FAINT_SOURCE
A.*=BRIGHT_COMPACT
\end{verbatim}
\end{quote}

would select \texttt{REDUCE\_SCIENCE} whenever a \emph{science}
observation of OBJECT1 is encountered but choose
\texttt{REDUCE\_FAINT\_SOURCE} whenever OBJECT2 is found. The third
line is an example of a regular expression that can be used to select
recipes based on a more general pattern match of the object name. This relies
on header translation functioning to find the observation type and
object name correctly. This sort of configuration is quite common when the
Observing Tool has not been set up to switch recipes.

Once a \recipe\ has been selected it can be configured as simple
key-value pairs:

\begin{quote}
\begin{verbatim}
[REDUCE_SCIENCE]
PARAM1 = value1
PARAM2 = value2

[REDUCE_SCIENCE:A.*]
PARAM1 = value3
\end{verbatim}
\end{quote}

and here, again, the parameters selected can be controlled by a
regular expression on the object name. The final set of parameters are
made available to the primitives in a key-value lookup table.

\subsection{Recipe Execution}
\label{sec:exec}

Once a set of files have been found the header is read to determine
how the data should be reduced. Files from the same observation are
read into what is known as a \Frame\ object. This object contains all
the metadata and pipeline context and, given that the currently used
applications require files to be written, the name of the
currently active intermediate file (or files for observations that
either consist of multiple files or which generate multiple
intermediate files). In some cases, such as for ACSIS, a single
observation can generate multiple files that are independent and in
these cases multiple \Frame\ objects are created and they are
processed independently. There is also a \Group\ object which
contains the collection of \Frame\ objects that the pipeline should
combine.

The pipeline will have been initialized to expect a particular instrument and
the resulting \Frame\ and \Group\ objects will be instrument-specific subclasses.

The \Frame\ object contains sufficient information to allow the
pipeline to work out which \recipe\ should be used to reduce the
data. The \recipe\ itself is located by looking through a search path
and modifiers can be specified to select recipe variants. For example,
if the recipe would normally be \texttt{REDUCE\_SCIENCE} the pipeline
can be configured to prefer a recipe suffix of \texttt{\_QL} to
enable a quick-look version of a recipe to be selected at the summit
whilst selecting the full recipe when running off-line.
The top-level \recipe\ is parsed and is then evaluated in the
parent pipeline context using the Perl \texttt{eval} function. The
\recipe\ is called with the relevant \Frame, and \Group\ objects along
with other context. The
reason we use \texttt{eval} rather than running the recipe in a
distinct process is to allow the recipe to update the state. As
discussed in \secref{sec:onvoff}, the pipeline is designed to
function in an incremental mode where data are reduced as they arrive,
with group co-adding either happening incrementally or waiting for a
set cadence to complete. This requires that the group processing stage
knows the current state of the \Group\ object and of the contributing
\Frame\ objects. Launching an external process to execute the
recipe each time new data arrived would significantly complicate the
architecture.

As noted in the previous section, the \recipe\ is parsed incrementally
and the decision on whether to re-read a \primitive\ is deferred until
that \primitive\ is required. This is important for instruments such
as MICHELLE and UIST which can observe in multiple modes
(spectroscopy, imaging, IFU), sometimes
requiring a single recipe invocation to call \primitives\ optimized
for the different modes. The execution environment handles this by
allowing a caller to set the instrument mode and this dynamically
adjusts the \primitive\ selection code.

\subsection{Header Translation}

As more instruments were added to \oracdr\ it quickly became apparent
that many of the \primitives\ were being adjusted to support different
variants of FITS headers through the use of repetitive if/then/else
constructs. This was making it harder to support the code and it was
decided to modify the \primitives\ to use standardized headers. When a
new \Frame\ object is created the headers are immediately translated
to standard form and both the original and translated headers are
available to \primitive\ authors.

The code to do the translation was felt to be fairly generic and was
written to be a standalone
module\footnote{\texttt{Astro::FITS::HdrTrans}, available on
  CPAN}. Each instrument header maps to a single translation class
with a class hierarchy that allows, for example, JCMT instruments to
inherit knowledge of shared JCMT headers without requiring that the
translations be duplicated. Each class is passed the input header and
reports whether the class can process it, and it is an error for multiple
classes to be able to process a single header. A method exists for each
target generic header where,
for example, the method to calculate the start airmass would be
\texttt{\_to\_AIRMASS\_START}. The simple unit mappings (where there
is a one-to-one mapping of an instrument header to a generic header
without requiring changes to units) are defined as simple Perl lookup tables
but at compile-time the corresponding methods are generated so that
there is no difference in interface for these cases. Complex mappings
that may involve multiple input FITS headers, are written as explicit
conversion methods.

The header translation system can also reverse the mapping such that a
set of generic headers can be converted back into instrument-specific
form. This can be particularly useful when required to update a header
during processing.

\subsection{Calibration System}

During \Frame\ processing it is necessary to make use of calibration
frames or parameters derived from calibration observations. The early
design focused entirely on how to solve the problem of selecting the
most suitable calibration frame for a particular science observation
without requiring the instrument scientist to write code or understand
the internals of the pipeline. The solution that was adopted involves
two distinct operations: filing calibration results and querying those results.
When a calibration image is reduced (using the same pipeline
environment as science frames) the results of the processing are
registered with the calibration system. Information such as the name
of the file, the wavelength, and the observing mode are all stored in the \Index.
In the current system the \Index\ is a text file on disk that is cached by
the pipeline but the design would be no different if an SQL database
was used instead; no \primitives\ would need to be modified to switch
to an SQL backend.  The only requirement is that the \Index\ is
persistent over pipeline restarts (which may happen a lot during
instrument commissioning).

The second half of the problem was to provide a rules-based system.
A calibration rule simply indicates how a header in the science data
must relate to a header in the calibration database in order for the
calibration to be flagged as suitable. The following is an excerpt
from a rules file for an imaging instrument dark calibration:

\begin{quote}
{\small
\begin{verbatim}
OBSTYPE eq 'DARK'
MODE eq $Hdr{MODE}
EXP_TIME == $Hdr{EXP_TIME}
MEANCOUNT
\end{verbatim}
}
\end{quote}

Each row in the rules file is evaluated in turn by replacing the
unadorned keyword with the corresponding calibration value read from
the \Index\ and the \texttt{\$Hdr} corresponding to the science
header. In the above example the
calibration would match if the exposure times and observing readout
mode match and the calibration itself is a dark.
These rules are evaluated using the Perl \texttt{eval} command
so the full Perl interpreter is available. This allows for
complex rules to be generated such as a rule that allows a calibration to expire
if it is too old.

The rules file itself represents the schema of the database in
that for every line in the rules file, information from that
calibration is stored in the \Index. In the example above,
\texttt{MEANCOUNT} is not used in the rules processing but the
presence of this item means that the corresponding value will be
extracted from the header of the calibration image and registered in
the calibration database. Once an item is stored in the calibration
database a calibration query will make that value available in
addition to the name of the matching calibration file.
It is therefore simple for the instrument
scientist to add a new header for tracking, although this does require
that the old \Index\ is removed and the data reprocessed to regenerate
a new \Index\ in the correct form.

The calibration selection system can behave differently in off-line
mode as the full set of calibrations can be made available and
calibrations taken after the current observation may be relevant. Each
instrument's calibration class can decide whether this is an
appropriate behavior.

The calibration system can also be modified by a command-line argument at
run time to allow the user to decide which behavior to use. For
example, with the SCUBA pipeline \citep{1999ASPC..172..171J} the user
can decide which opacity calibration scheme they require from a number
of options.

One of the more controversial aspects of the calibration system was
that the UKIRT pipelines would stop and refuse to reduce data if no
suitable calibration frame had been taken previously (such as a dark
taken in the wrong mode or with the wrong exposure). This sometimes
led to people reporting that the pipeline had crashed (and so was
unstable) but the purpose was to force the observer to stop and think
about their observing run and ensure that they did not take many hours
of data with their calibration observations being taken in a manner
incompatible with the science data. A pro-active pipeline helped to
prevent this and also made it easier to support flexible scheduling
\citep{2002ASPC..281..488E,2004SPIE.5493...24A} without fearing that
the data were unreducible.

This hard-line approach to requiring fully calibrated observations,
even if the PI's specific science goals did not require it, was
adopted in anticipation of the emergence of science data archives as
an important source of data for scientific papers. Casting the PI not
as the data owner, but rather as somebody who is being leased
observatory data from the public domain for the length of their
proprietary period, requires an observation as only being complete if
fully calibratable. In that way, the telescope time's value is
maximised by making the dataset useful to the widest range of its
potential uses. To this end, the authors favor a model where for
flexibly-scheduled PI-led facilities, calibration time is not deducted
from the PI's allocation.

\subsection{Provenance Tracking}
\label{sec:prov}

For the outputs from a data reduction pipeline it is important for
astronomers to understand what was done to the data and how they can
reproduce the processing steps. \oracdr\ manages this provenance and
history tracking in a number of different ways. The pipeline makes available to
\primitives\ the commit ID (SHA1) of the pipeline software and the
commit ID of the external application package. It is up to the
\recipe\ to determine whether use should be made of that
information. For the \recipes\ that run at the JCMT Science Archive
\citep{2014Economou} there is code that inserts this information, and
the \recipe\ name, into data headers. Summit processing \recipes\ do
not include this detail as the products are generally thought to be
transient in nature as the \recipes\ are optimized for speed and
quality assurance tracking rather than absolute data quality. One
caveat in this approach is that an end-user who modifies a
\recipe\ will not see any change as the commit ID will not have
changed. This was thought to be of secondary importance compared to
the major use case of archive processing but does need consideration
before the reproducibility aspects of data reduction can be considered
complete.

Detailed tracking of the individual steps of the processing are
handled differently in that the pipeline is written with the
assumption that the external applications will track provenance and
history themselves. This is true for the Starlink software where the
NDF library, which already supported detailed history tracking, was
updated to also support file provenance so that all ancestor files
could be tracked \citep[see e.g.][for details on the provenance algorithm]{ndfjenness}.
We took this approach because we felt it was far too complicated to
require that the pipeline infrastructure and \primitives\ track what
is being done to the data files. Modifying the file I/O library meant
that provenance tracking would be available to all users of the
external packages (in this case the Starlink software applications)
and not just the pipeline users. The history information automatically logged by the external
applications is augmented by code in the pipeline that logs the
primitive name whenever header information is synchronized to a file,
and, optionally, all text messages that are output by a \primitive\ can be
stored as history items in the files written by the \primitive.

\subsection{Configurable Display System}

On-line pipelines are most useful when results are displayed to the
observer. One complication with pipeline display is that different
observers are interested in different intermediate data products or
wish the final data products to be displayed in a particular
way. Display logic such as this can not be embedded directly in
\primitives; all a \primitive\ can do is indicate that a particular
product \emph{could} be displayed and leave it to a different system
to decide \emph{whether} the product should be displayed and how to
do so.

The display system uses the \oracdr\ file naming convention to
determine relevance. Usually, the text after the last underscore,
referred to as the file suffix, is used to indicate the reduction step
that generated the file: \texttt{mos} for mosaic, \texttt{dk} for
dark, etc. When a \Frame\ or \Group\ is passed to the display system
the file suffix and, optionally a \Group\ versus \Frame\ indicator,
are used to form an identifier which is compared with the entries in
the display configuration file. For each row
containing a matching identifier the files will be passed to the
specific display tool. Different plot types are available such as
image, spectrum, histogram, and vector plot and also a specific mode
for plotting a 1-dimensional dataset over a corresponding model. Additional
parameters can be used to control placement within a viewport and how
auto-scaling is handled. The display system currently supports \textsc{gaia}
\citep[][\ascl{1403.024}]{2009ASPC..411..575D} and \textsc{kappa}
\citep[][\ascl{1403.022}]{SUN95} as well as the historical P4 tool
(part of \cgsdr\ \citep{SUN27} and an important influence on the
design).

Originally the display commands would be handled within the \recipe\
execution environment and would block the processing until the display
was complete. This can take a non-negligible amount of time and for the
SCUBA-2 pipeline to meet its performance goals this delay was
unacceptable. The architecture was therefore modified to allow the
display system running from within the \recipe\ to register the
display request but for a separate process to be monitoring these
requests and triggering the display.

\subsection{Support modules}

As well as the systems described above there are general support
modules that provides standardized interfaces for message output, log files
creation and temporary file handling.

The message output layer is required
to allow messages from the external packages and from the \primitives\
to be sent to the right location. This might be a GUI, the terminal or
a log file (or all at once) and supports different messaging levels to
distinguish verbose messages, from normal messages and
warnings. Internally this is implemented as a tied object that
emulates the file handle API and contains multiple objects to allow
messages to be sent to multiple locations.

Log files are a standard requirement for storing information of
interest to the scientist about the processing such as
quality assurance parameters or photometry results. The pipeline
controls the opening of these files in a standard way so that the
primitive writer simply has to worry about he content.

With the current external applications there are many intermediate files
and most of them are temporary. The allocation of filenames is handled
by the infrastructure and they are cleaned up automatically unless the
pipeline is configured in debugging mode to retain them.

\section{Supporting New Instruments}

An important part of the \oracdr\ philosophy is to make adding new
instruments as painless as possible and re-use as much of the
existing code as possible. The work required obviously depends on the
type of instrument. An infrared array will be straightforward as many
of the \recipes\ will work with only minor adjustments. Adding support
for an X-Ray telescope or radio interferometer would require
significantly more work on the recipes.

To add a new instrument the following items must be considered:

\begin{itemize}

\item How are new data presented to the pipeline? \oracdr\ supports a
  number of different data detection schemes but can't cover every option.

\item What is the file format? All the current \recipes\ use Starlink
  applications that require NDF \citep{ndfjenness} and if FITS
  files are detected the infrastructure converts them to NDF before
  handing them to the rest of the system. If the raw data are in HDF5,
  or use a very complex data model on top of FITS, new code will have
  to be written to support this.

\item How to map the metadata to the internal expectations of the
  pipeline? A new module would be needed for \texttt{Astro::FITS::HdrTrans}.

\item Does it need new \recipes/\primitives? This depends on how close
  the instrument is to an instrument already supported. The \recipe\
  parser can be configured to search in instrument-specific
  sub-directories and, for example, the Las Cumbres Observatory
  imaging recipes use the standard \primitives\ in many case but also
  provide bespoke versions that handle the idiosyncrasies of their
  instrumentation.

\end{itemize}

Once this has been decided new subclasses will have to be written to
encode specialist behavior for \Frame\ and \Group\ objects and the
calibration system, along with the instrument initialization class
that declares the supported calibrations and applications.

\section{Lessons Learned}

\subsection{Language choice can hinder adoption}

In 1998 the best choice of dynamic ``scripting'' language for an astronomy project was
still an open question with the main choices being between Perl and
Tcl/Tk with Python being a distant third
\citep{1995ComPh...9...57A,1999ASPC..172..494J,1999ASPC..172..483B,2000ASPC..216...91J}.
Tcl/Tk had already been adopted by Starlink
\citep{1995ASPC...77..395T}, STScI \citep{1998SPIE.3349...89D},
SDSS \citep{1996ASPC..101..248S} and ESO \citep{1996ASPC..101..396H,1995ASPC...77...58C} and
would have been the safest choice, but at the time it was felt that
the popularity of Tcl/Tk was peaking. Perl was chosen as it was a language
gaining in popularity and the development team were proficient in
it in addition to developing the Perl Data Language \citep[PDL;][]{PDL}
promising easy handling of array data; something Tcl/Tk was incapable
of handling.

Over the next decade and a half, beginning with the advent of \texttt{pyraf}
\citep[][\ascl{1207.010}]{2000ASPC..216...59G,2006hstc.conf..437G}
and culminating in Astropy \citep[][\ascl{1304.002}]{2013A&A...558A..33A},
Python became the dominant language for astronomy,
becoming the \emph{lingua franca} for new students in astronomy and
the default scripting interface for new data reductions systems such
as those for ALMA
\citep{2007ASPC..376..127M} and LSST \citep{2010SPIE.7740E..15A}.
In this environment, whilst \oracdr\ received much interest from other
observatories, the use of Perl rather than Python became a
deal-breaker given the skill sets of development groups. During this
period only two additional observatories adopted the pipeline: the
Anglo-Australian Observatory for IRIS2 \citep{2004SPIE.5492..998T} and Las Cumbres
Observatory for their imaging pipeline \citep{2013PASP..125.1031B}.

The core design concepts were not at issue, indeed, Gemini adopted the
key features of the \oracdr\ design in their Gemini Recipe System
\citep{2014ASPC..485..359L}. With approximately 100,000 lines of Perl code in
\oracdr\footnote{For infrastructure and \primitives, but counting code only, with comments adding more than
  100,00 lines to that
  number. Blank line count not included, nor are support modules from CPAN
  required by the pipeline but distributed separately.} it
is impractical to rewrite it all in Python given that the system does
work as designed.

Of course, a language must be chosen without the benefit of hindsight
but it is instructive to see how the best choice for a particular
moment can have significant consequences 15 years later.

\subsection{In-memory versus intermediate files}

When \oracdr\ was being designed the choice was between IRAF
\citep[][\ascl{9911.002}]{2012ASPC..461..595F} and Starlink for the
external packages.
At the time the answer was that Starlink messaging and error reporting were
significantly more robust and allowed the \primitives\ to adjust their
processing based on specific error states (such as there being too few
stars in the field to solve the mosaicking offsets). Additionally,
Starlink supported variance propagation and a structured data format.
From a software
engineering perspective Starlink was clearly the correct choice but it
turned out to be yet another reason why \oracdr\ could not be adopted
by other telescopes. Both these environments relied on each command
reading data from a disk file, processing it in some way and then
writing the results out to either the same or a new file. Many of
these routines were optimized for environments where the science data
was comparable in size to the available RAM and went to great lengths
to read the data in chunks to minimize swapping. It was also not
feasible to rewrite these algorithms (that had been well-tested) in
the Perl Data Language, or even turn the low-level libraries into Perl
function calls, and the penalty involved in continually reading
and writing to the disk was deemed to be a good trade off.

As it turns out, the entire debate of Starlink versus IRAF is somewhat
moot in the current funding climate and in an era where many pipeline
environments \citep[e.g.,][]{2010SPIE.7740E..15A} are abandoning
intermediate files and doing all processing in memory for performance
reasons, using, for example, \texttt{numpy} arrays or ``piddles''\footnote{A
 ``piddle'' is the common term for an array object in the Perl Data Language;
 an instance of a \texttt{PDL} object.}. For
instruments where the size of a single observation approaches 1\,TB
\citep[e.g., SWCam at CCAT;][]{2014SPIE9153-21} this presents a
sizable challenge but it seems clear that this is the current trend
and a newly written pipeline infrastructure would assume that all
algorithm work would be in memory.

\subsection{Recipe configuration is needed}

Initially, the intent was for \recipes\ to be edited to suit different
processing needs and for the processing to be entirely driven by the
input data. This was driven strongly by the requirement that the
pipeline should work at the telescope without requiring intervention
from the observer. The initial design was meant to be that the
astronomer would select their \recipe\ when they prepared the
observation and that this would be the \recipe\ automatically picked
up by the pipeline when the data were observed. Eventually we realized
that anything more than two or three recipes to choose from (for
example, is your object broad line or narrow line, or are your objects
extremely faint point sources or bright extended structures?) in the
Observing Tool became unwieldy and most people weren't sure how they
wanted to optimize their data processing until they saw what the
initial processing gave them.

After many years of resistance a system was developed in 2009 for
passing \recipe\ parameters from configuration files to the pipeline
and this proved to be immensely popular. It is much simpler for people
to tweak a small set of documented parameters than it is to edit
recipes and it is also much easier to support many project-specific
configuration files than it is to keep track of the differences
between the equivalent number of bespoke recipes. When a processing
job is submitted to the JCMT Science Archive any associated
project-specific configuration file is automatically included, and
these can be updated at any time based on feedback from the data
products. It took far too long to add this functionality and this
delay was partly driven by the overt focus on online functionality
despite the shift to the pipeline being used predominantly in an
offline setting. This is discussed further in the next section.

\subsection{Online design confused offline use}
\label{sec:onvoff}

\oracdr\ was initially designed for on-line summit usage where data
appear incrementally and where as much processing should be done on
each frame whilst waiting for subsequent frames to arrive. As
discussed previously (\secref{sec:exec}), this led to the
execution of the \recipe\ within the main process so that context
could be shared easily.

For off-line mode the environment is very different and you would
ideally wish to first reduce all the calibration observations, then
process all the individual science observations and finally do the
group processing to generate mosaics and co-adds. When doing this the
only context that would need to be passed between different \recipe\
executions would be the calibration information that is already
persistent. Indeed, the \recipes\ themselves could be significantly
simplified in that single observation \recipes\ would not include any
group processing instructions. This is not strictly possible in all
cases. For the ACSIS data reduction \recipes\ \citep{JennessACSISDR}
the output of the frame processing depends on how well the group
co-adding has been done; the more measurements that are included, the
better the baseline subtraction.

As written, the recipes have to handle both on-line and off-line
operation and this is achieved by the group \primitives\ being
configured to be no-ops if they realize that the \Frame\ object that
is currently being processed is not the final member of the group.
Whilst the off-line restrictions can be annoying to someone reducing a
night of data on their home machine, it is possible to deal with the
problems by scanning through a set of observations and running the
pipeline separately for each one. This is exactly how the the healpix
processing for the JCMT Science Archive is implemented
\citep{2014SPIE9152-93}. Parallelization is therefore occurring at a
level above the \oracdr\ pipeline itself. Currently, the UKIRT Science
Archive \citep{2014ASPC..485..143B} reduces data in two passes: first
all the calibrations are reduced from a single night and then all the
science data are reduced together. This is particularly important for
the archival data taken before the ORAC software was released.
In 2012 Las Cumbres added the ability to sort the \Group\ objects
before starting the processing in off-line mode and this allows them to
reduce all the calibration observations before doing the science
observations.

The PiCARD \citep{SUN265} frontend to the infrastructure was developed
to try to overcome some of the on-line bias in the data handling. The
\textbf{Pi}peline for \textbf{C}ombining and \textbf{A}nalyzing
\textbf{R}educed \textbf{D}ata, was introduced in 2007
\citep{2008ASPC..394..565J} specifically to allow the existing
infrastructure to be leveraged in an off-line science archive
environment. The \oracdr\ layer was removed and replaced with a
simplified application that accepts a list of files, works out what
type of instrument they come from, and uses the same parser to read
the specified recipe.

\subsection{Dynamic recipe generation works well}

Whilst the initial driver for dynamic recipe generation came directly
out of the requirement for a readable \recipe, it would have been
simple to write the \primitives\ as Perl subroutines from the
beginning and require the primitive writer to handle the subroutine
arguments and return codes. In some sense, this would have been the
obvious approach as that is how most people want to write code.

It soon became clear that the benefits associated
with running a parser over the \primitives\ were substantial. Not only
could we minimize the use of repetitive code but we could dynamically
add in profiling code. In fact, this was critical to the ongoing
development of the \oracdr\ infrastructure as the current version of
the parser is the third complete rewrite and none of the rewrites
required any change to \primitive\ code. The first version of the
parser did not use subroutines at all and simply expanded the \recipe\
into one long string for evaluation.

\subsection{Threading and Distributed Processing}

Perl does not really support multi-threaded
operation\footnote{There have been multiple attempts to support native
  threads in Perl but nothing has appeared that is recommended.}
and this has led to some problems with components of the system that
use blocking I/O such as the display system or the file detection
code. Being able to monitor new data arriving whilst creating a thread for the
current data to be processed seems attractive but would probably have
created more complication than would be acceptable and using separate
processes has worked despite it feeling ``kludgy''.

Similarly, the way the messaging interface was implemented meant that
it was not possible to send multiple messages concurrently. The
ability to send four files to four separate, but identical, tasks in
parallel, and waiting for them all to complete would have led to some
efficiency games even if the tasks were all running on the single
pipeline computer. We are again struggling with this issue as
the \textsc{smurf} map-maker \citep[][\ascl{1310.007}]{2013MNRAS.430.2545C}, currently supported by
\oracdr, is being modified to support
multi-node operation using MPI \citep{2014ASPC..485..399M}. It may be
that implementing MPI at the task level is much easier than at the
messaging level but this remains to be tested.

\section{Conclusions}

\oracdr\ has been in use since 1998, running on four telescopes and ten
instruments covering the submillimeter to the optical and imaging to
IFUs. Prototype pipelines have also been developed for Gemini and ESO
instruments to prove the inherent flexibility of the design
\citep{2003ASPC..295..237C,2004ASPC..314..460C} and the CCAT project
are considering adopting it \citep{2014SPIE9152-109}.

In an environment similar to that now promoted by the DevOps movement
\citep[see e.g.,][]{2014arXiv1407.6463E},
we participated, together with our contributors, in the design,
implementation, extension, operation, distribution and night-time
telescope support of the pipeline as it was running on every night at
UKIRT, JCMT and later the JCMT Science Archive over a period of 16
years. We therefore had the opportunity to evaluate what features
contributed to the easy extensibility and low maintenance cost of the
codebase over a long period of time in a changing computing environment.

We regard the following as key design successes of \oracdr, and hope
to see them repeated in other astronomical data reduction
infrastructures:

\begin{itemize}

\item The clean separation of recipes, primitives, instrument
  specification, logging, and external applications. The core design is
  unchanged from the beginning and we frequently reaped the benefits
  of a design that meant that the science code was well isolated from
  the infrastructure components that were in turn relatively agnostic
  about their implementation. Thus we were able to engage in major
  pieces of refactoring such as rewriting the recipe parser or
  reorganizing the class hierarchy for \Frame\ and \Group\ classes
  without affecting the pipeline users.

\item Reusing primitive code for new instruments was a key requirement
  and this has worked; especially once header translation was
  added. For example, Las Cumbres Observatory are adding additional
  instruments to the pipeline and can reuse the imaging \primitives,
  only modifying code that defines the differences between
  instruments. A similar approach has been used by the LSST data
  management stack to allow it to support different camera geometries
  \citep{2010SPIE.7740E..15A}.

\item Defining the items of interest in the calibration rules file
  guarantees that all the required information is available to the
  calibration system whilst also making it trivial to add additional
  information to the database. Such techniques anticipate the evolving
  requirements of astronomical observing and build in a flexibility
  that requires no, or minimal, changes. It also externalizes to
  instrument scientists important aspects of the pipeline operations
  in a human readable (and configurable) way.

\item \oracdr\ was used at UKIRT and JCMT as a ``quick-look'' pipeline
  during observing (even though the data reduction quality was higher
  than the term suggests) and so was a vital part to the smooth
  functioning of the observatory. As is often the case, the pipeline
  was often the first system to alert the observers that something had
  gone awry with data taking, and so occasionally triggered night-time
  support calls reporting that ``there was something wrong with the
  pipeline'', despite the fact that an extensive suite of engineering
  interfaces were also available to the operators. We conclude from
  this that despite the understandable tendency to focus on the
  scientific user of the data, data reduction systems are often the
  first port of call for investigating technical issues. We frequently
  utilized the completeness and readability of our file metadata to
  identify problems for our engineers, and so came to regard the data
  reduction infrastructure as a core technical element of a modern
  facility as well as a valuable scientific productivity tool.

\end{itemize}

Ultimately, the design of the pipeline infrastructure succeeded in its
goals of minimizing the amount of code needed for each new instrument
delivery, re-using a significant amount of code available to the
community and in making a system that can be tweaked easily by
instrument scientists and interested astronomers.

\section*{Acknowledgments}

\oracdr\ was initially developed for UKIRT as part of the ORAC Project
\citep{1998SPIE.3349..184B}, and then further developed by the Joint
Astronomy Centre to support JCMT and UKIRT instrumentation. We thank
all the \oracdr\ recipe developers who have helped make the pipeline
successful, and in particular we thank  Malcolm Currie, Brad Cavanagh,
Andy Gibb, Paul Hirst, Tim Lister, Stuart Ryder and Stephen Todd for their
significant contributions. We thank Malcolm Currie and Andy Gibb for
comments on the draft manuscript.

The source
code is available from
\htmladdnormallinkfoot{Github}{https://github.com/Starlink/ORAC-DR}
and distributed under the Gnu General Public Licence.

\end{document}